\documentclass[12pt,letterpaper]{article}
\pdfoutput=1 % if your are submitting a pdflatex (i.e. if you have
\date{}
\input{epsf}
\usepackage{jheppub}
\usepackage[all]{xy}
\usepackage{amsmath,amssymb,amsfonts,amsxtra,amscd,mathrsfs,graphics,graphicx,amsthm,epsfig, youngtab,bm,longtable,float,tikz,empheq}
\usepackage[margin=1cm]{caption}
\usepackage{lipsum}
\usepackage{ragged2e}
\usepackage{slashed}
\usetikzlibrary{positioning}

\newcommand{\ba}{\begin{array}}
\newcommand{\ea}{\end{array}}
\newcommand{\bi}{\begin{itemize}}
\newcommand{\ei}{\end{itemize}}
\newcommand{\de}{{\rm d}}

\def\bea#1\eea{\allowdisplaybreaMs \begin{align}#1\end{align}}
 \newcommand{\ben}{\begin{enumerate}}
\newcommand{\een}{\end{enumerate}}
\newcommand{\bean}{\begin{eqnarray*}}
\newcommand{\eean}{\end{eqnarray*}}
\newcommand{\eref}[1]{(\ref{#1})}

\newcommand{\comment}[1]{}

\newcommand{\CO}{{\cal O}}

\newcommand{\CN}{{\cal N}}

\newcommand{\I}{{\mathrm i}}

\newcommand{\Secref}[1]{Section~\ref{#1}}
\newcommand{\secref}[1]{Sec.~\ref{#1}}

\newcommand{\figref}[1]{Fig.~\ref{#1}}
\renewcommand{\eqref}[1]{(\ref{#1})}

\title{Good IR Duals of Bad Quiver Theories}
\author[a]{Anindya Dey,}
\author[b]{Peter Koroteev}

\affiliation[a]{New High Energy Theory Center, Rutgers University, Piscataway, NJ 08854, USA}
\affiliation[b]{Department of Mathematics, University of California, Davis, CA 95616, USA}
\affiliation[b]{Kavli Institute for Theoretical Physics, University of California, Santa Barbara, CA  93106, USA}
\emailAdd{anindya.hepth@gmail.com}
\emailAdd{pkoroteev@math.ucdavis.edu}

\abstract{The infrared dynamics of generic 3d $\CN=4$ bad theories (as per the good-bad-ugly classification of Gaiotto and Witten) are 
poorly understood. Examples of such theories with a single unitary gauge group and fundamental flavors have been studied recently,
and the low energy effective theory around some special point in the Coulomb branch was shown to have a description in terms of a 
good theory and a certain number of free hypermultiplets. A classification of possible infrared fixed points for bad theories by Bashkirov, 
based on unitarity constraints and superconformal symmetry, suggest a much richer set of possibilities for the IR behavior, although explicit 
examples were not known. In this note, we present a specific example of a bad quiver gauge theory which admits a good IR description on a 
sublocus of its Coulomb branch. The good description, in question, consists of two decoupled quiver gauge theories with no free hypermultiplets.}

\preprint{NSF-ITP-17-150}

\begin{document}
\maketitle

\section{Generalities and Summary of Results}\label{Sec:Bad}
The notion of {\it good, bad} and {\it ugly} theories was introduced by Gaiotto and Witten \cite{Gaiotto:2008ak} in order to classify $3d$ $\CN=4$ 
theories according to their expected IR behavior. A bad theory is defined as one for which the IR scaling dimensions of certain BPS monopole 
operators violate the unitarity bound. Such a theory cannot flow in the IR to an SCFT whose R-symmetry matches with the R-symmetry 
of the UV description, and therefore the IR behavior of these theories require more careful treatment. Good theories, on the other hand, 
have monopole operators which strictly obey the unitarity bound, and therefore flow to an IR SCFT whose R-symmetry can be directly read off 
from its UV description. Ugly theories have some monopole operators which saturate the unitarity bound and flow in the IR to a standard SCFT (i.e one 
whose R-symmetry is visible in the UV) plus some additional free hypermultiplets.\\
\indent In particular, for a $U(N_c)$ theory with $N_f$ flavors, one encounters a bad theory for $N_f < 2N_c -1$, while the condition for good and ugly theories are $N_f \geq 2N_c$ and $N_f = 2N_c -1$ respectively. In \cite{Yaakov:2013fza}, certain Seiberg-like good duals were proposed for these bad theories with $N_c \leq N_f < 2N_c -1$ using sphere partition functions. For $N_c = N_f$, the proposed dual is a collection of $N_f$ free twisted hypermultiplets, while for $N_c < N_f < 2N_c -1$ the proposed dual is a $U(N_f - N_c)$ theory with $N_f$ flavors. This claim was confirmed in \cite{Gaiotto:2013bwa} by studying spaces of supersymmetric vacua of $\CN=2^*$ theories with the same gauge and matter content as well as in \cite{Hwang:2015wna,Hwang:2017kmk} by working with superconformal index and vortex partition functions. \\
\indent Recently, careful algebraic analysis of \cite{Assel:2017jgo} motivated by \cite{Braverman:2016wma,Nakajima:2015txa,Bullimore:2015lsa} showed that 
the proposed duality is {\it not} an exact duality: the moduli spaces of the proposed duals are different globally. The good `dual' theory should be thought of 
as the correct low energy description at a very special point on the moduli space of the bad theory. However, turning on an Fayet-Illiopoulos parameter for the $U(1)$ factor 
of the gauge group lifts the Coulomb branch of the bad theory aside from this special point, and in this case the proposed duality becomes exact.\\

IR dynamics of bad theories analyzed in \cite{Yaakov:2013fza, Assel:2017jgo} involve a single interacting SCFT and free hypermultiplets.
Fairly general considerations based on the superconformal algebra and unitarity constraints led the author of \cite{Bashkirov:2013dda} to 
propose the following classification of IR SCFTs which can be realized as IR fixed points of UV bad theories: 
\begin{enumerate}
\item Interacting $\CN=4$ SCFT whose flavor symmetry group $G$ has several $SU(2)$ subgroups, i.e. $SU(2)^k\subset G$ for some integer $k$.
\item Irreducible $\CN=8$ SCFT.
\item Free $\CN=4$ hypermultiplets.
\item Union of free hypermultiplets and/or interacting $\CN=4$ SCFTs, in particular, all sectors may be interacting.
\end{enumerate}
Aside from the example described above, very little is known about the IR dynamics of bad theories. 
In particular, explicit examples of bad theories which flow in the IR to a set of decoupled interacting SCFTs are not known.
In this paper, we present a specific example of a bad theory whose IR description belongs to class $4.$ above -- it consists of
two decoupled interacting SCFTs each of which has a quiver description in the UV, with no free hypermultiplets. A more systematic 
analysis of bad quiver gauge theories will be the subject of a future work.\\

The main results of this paper are summarized as follows:
\begin{itemize}
\item The bad theory and the proposed good dual\footnote{We would like to emphasize that the moduli spaces of the bad and the proposed good theory, defined on a flat space-time, are not isomorphic. Instead, there is a special sublocus on the moduli space of the bad theory, where the low energy effective field theory factorizes into an SCFT 
(U(1) with two flavors with masses and FI parameters tuned to zero) and the Coulomb branch of an SU(2) gauge theory with $N$ flavors. 
This is the sense in which we call the theories dual. } are given in \figref{fig: ex1}.  
The good dual consists of two decoupled quiver gauge theories on the right of the figure. 
 The duality holds only when the masses and FI parameter of the U(1) theory with two flavors 
are set to zero. The integer $N$ does not play a very important role in the analysis, and we will 
take $N=4$ for most of the paper to simplify the computation.

\begin{figure}[htbp]
\begin{center}
\begin{tikzpicture}[%node distance=1.3cm,
cnode/.style={circle,draw,thick, minimum size=1.5cm},snode/.style={rectangle,draw,thick,minimum size=1cm}]
\node[cnode] (1) {$SU(2)$};
\node[cnode] (2) [right=1cm  of 1]{$SU(2)$};
\node[snode] (3) [below=1cm of 1]{$SO(2N)$};
%\node[text width=0.1cm](20)[below=1cm of 3]{(a)};
%\node[snode] (19) [below=0.5cm of 9]{1};
%\node[text width=6cm](20)[above=0.5cm of 17]{$B=(1,0)$};
\draw[-] (1) -- (2);
\draw[-] (1)-- (3);
\end{tikzpicture}
%\caption{Higgs branch quiver for $\cM(B, \mathbf{v})$ in a $SU(2)$ theory for $B=(-p, p)$ and $\mathbf{v}=(-v, v)$.}
%\label{quiverU(2)}
%\end{figure}
%\end{center}
\qquad \qquad \qquad \qquad 
\begin{tikzpicture}[%node distance=1.3cm,
cnode/.style={circle,draw,thick, minimum size=1.5cm},snode/.style={rectangle,draw,thick,minimum size=1cm}]
\node[cnode] (1) {1};
\node[snode] (2) [below=1cm of 1]{2};
%\node[text width=0.1cm](20)[below=1cm of 3]{(b1)};
\draw[-] (1) -- (2);
\end{tikzpicture}
\qquad
\begin{tikzpicture}[%node distance=1.3cm,
cnode/.style={circle,draw,thick, minimum size=1.5cm},snode/.style={rectangle,draw,thick,minimum size=1cm}]
\node[cnode] (1) {$SU(2)$};
\node[snode] (2) [below=1cm of 1]{$SO(2N)$};
%\node[text width=0.1cm](20)[below=1cm of 3]{(b2)};
\draw[-] (1) -- (2);
\end{tikzpicture}
\caption{The quiver on the left-hand side is a bad theory. We claim that the quiver on the right-hand side -- constituted of two decoupled quivers -- is a good realization of the bad theory on the left.}
\label{fig: ex1}
\end{center}
\end{figure}
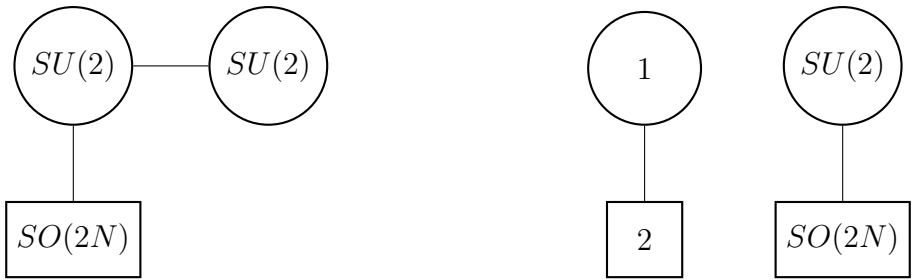

\item We present evidence for the proposed duality in \figref{fig: ex1} from three-dimensional mirror symmetry in 
\Secref{3dMS}.
In a Type IIB brane construction \cite{Hanany:1999sj}, the bad quiver arises as the mirror of an affine $D_4$ 
quiver with fundamental matter (as shown in \figref{fig:QuiverflavoredD44}). Localization computation, on the other
hand, leads to a good mirror of the affine $D_4$ quiver -- the quivers on the right in \figref{fig: ex1}. This suggests that
these quivers might constitute a good dual of the bad quiver. 
\begin{figure}[htbp]
\begin{center}
\includegraphics[scale=0.3]{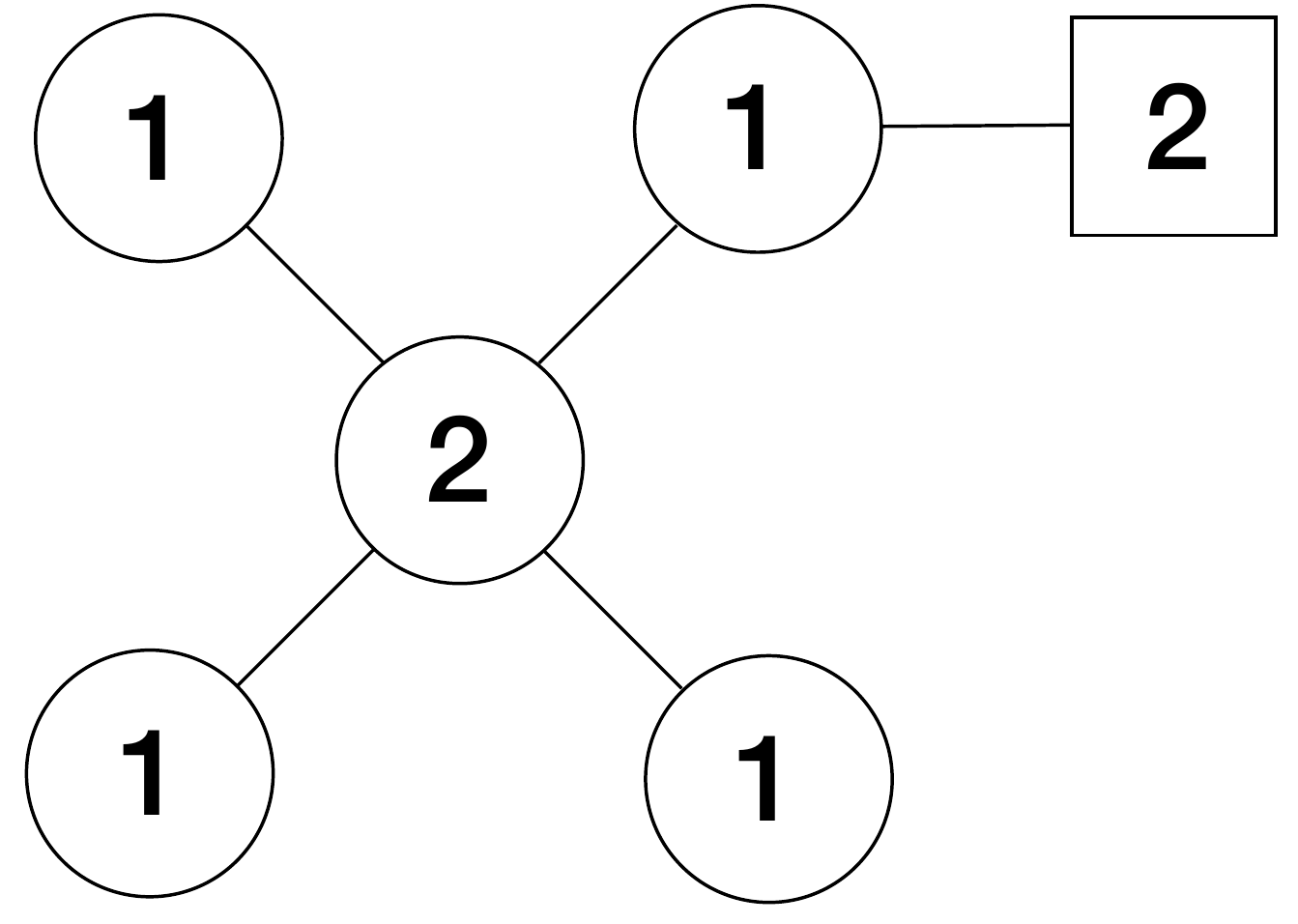}
\caption{Doubly framed $\widehat{D}_4$ quiver.} 
\label{fig:QuiverflavoredD44}
\end{center}
\end{figure}

\item In \Secref{Sec:CoulmbBranch}, we study the Coulomb branch of the bad quiver as an algebraic variety along the lines of \cite{Bullimore:2015lsa,Assel:2017jgo} and show that there exists a sublocus of the moduli space for which the good dual description is valid. We explicitly demonstrate by choosing the proper coordinates that the Coulomb branch of the bad theory factorizes into two components on this sublocus. 
%We match the R-symmetry of the bad theory with the R-symmetry of the IR SCFT.

\item In \Secref{PFA} we show how the good description arises from an appropriately regularized partition function on a three-sphere.

\end{itemize}

\section{Evidence from 3d Mirror Symmetry}\label{3dMS}
\begin{figure}[htbp]
\begin{center}
\includegraphics[scale=0.3]{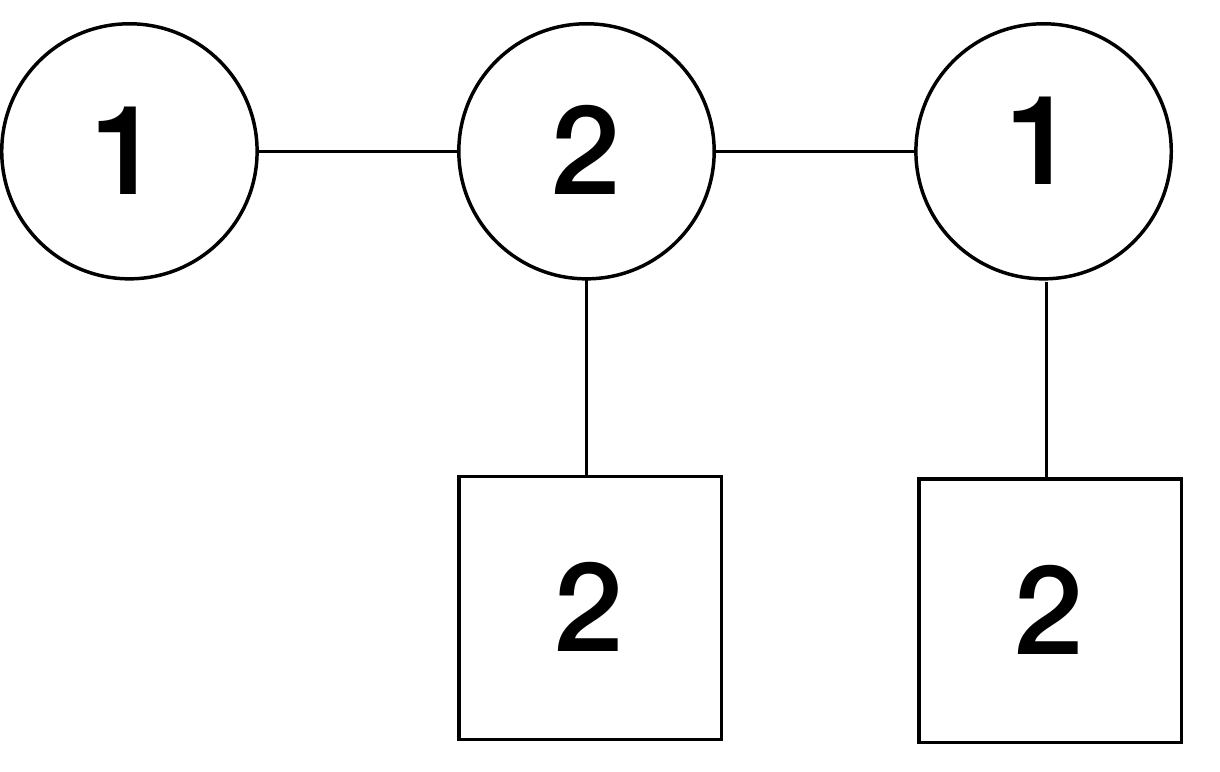}\qquad\qquad\qquad\includegraphics[scale=0.3]{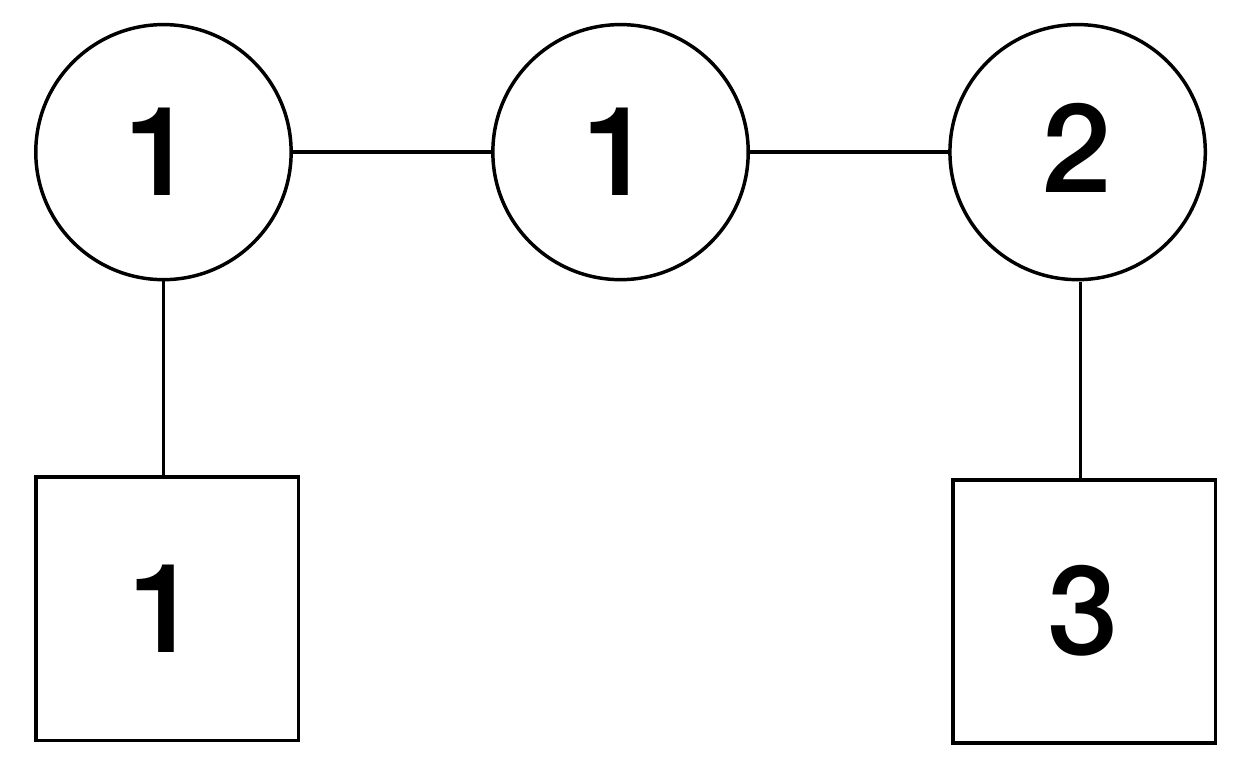}\\
\includegraphics[scale=0.55]{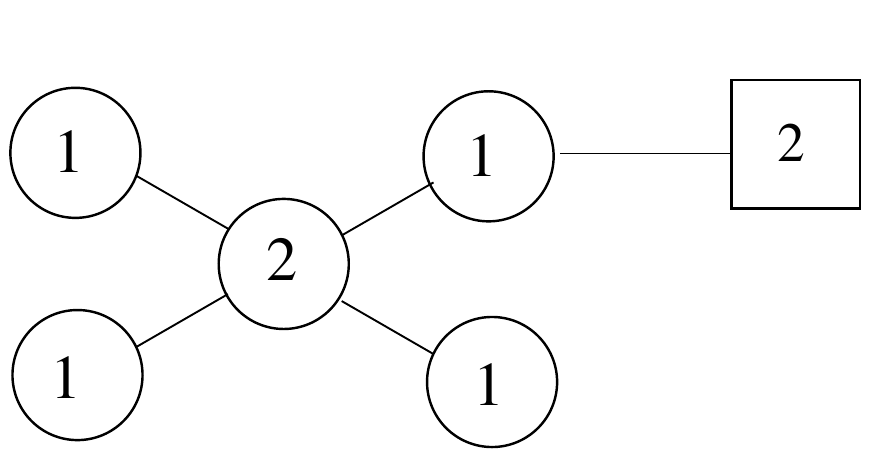}\qquad\qquad\qquad\includegraphics[scale=0.3]{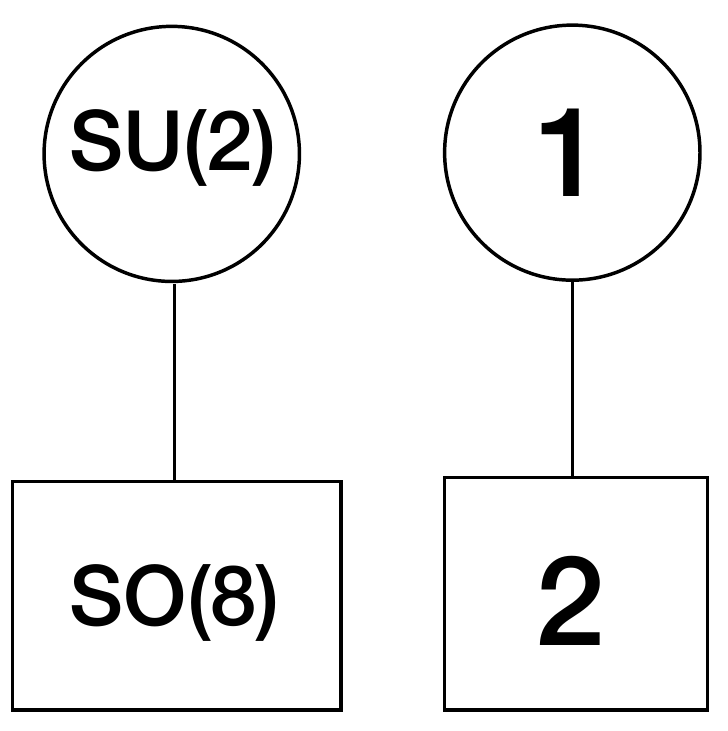}
\caption{Mirror dual framed linear quivers (top row) and the new mirror pair after applying gauging procedure.}
\label{fig:QuiverflavoredD44Lin}
\end{center}
\end{figure}

It was shown in \cite{Hanany:1999sj} using a Type IIB brane construction with orientifold planes that the quiver on the left in \figref{fig: ex1} appeared as the three-dimensional mirror dual of the quiver theory in \figref{fig:QuiverflavoredD44} (where we take $N=4$ for concreteness), which is an affine $\widehat{D}_4$ quiver 
with framing at one of the external nodes.  A good mirror of the latter quiver theory was derived in \cite{Dey:2014tka} using three-sphere partition functions. 
We give a very brief review of the result here and refer the reader to section of \cite{Dey:2014tka} for details. \\
\indent The starting point of the derivation is the pair of linear quivers in the top row of \figref{fig:QuiverflavoredD44Lin} which are mirror dual to each other.
Both theories are good quivers and have straightforward realization in terms of a D3-D5-NS5-brane system \cite{Hanany:1996ie}. In the next step, we 
gauge the $U(2)$ flavor symmetry on the middle $U(2)$ gauge node of the top left quiver as a $U(1) \times U(1)$ gauge group which leads to the affine $D_4$ 
quiver in the lower left corner. On the dual side, this translates to  an `ungauging'  operation on the top right quiver which breaks the $A_3$ linear quiver into two $A_1$ quivers, as shown in the lower right of \figref{fig:QuiverflavoredD44Lin}.

Therefore gauging/ungauging of the auxiliary mirror pair of quiver theories yields a good mirror directly, bypassing the bad quiver in \figref{fig:QuiverflavoredD44}. 
This naturally suggests that the quiver in the bottom left corner of \figref{fig:QuiverflavoredD44Lin} is a good dual of the bad theory in \figref{fig: ex1}. As an  
additional check, one can compare the Higgs branch Hilbert Series of the two theories which count the half-BPS chiral operators built out of hypermultiplets. This was 
done in section 4 of \cite{Dey:2014tka} and they were shown to agree. It was pointed out in \cite{Mekareeya:2015} that the splitting of the moduli space into a product of two hyper K\"ahler spaces is a general feature when a flavor node is attached to the affine node of any extended Dynkin diagram.\\

\section{Coulomb Branch Analysis}\label{Sec:CoulmbBranch}
Consider the theory on the left in \figref{fig: ex1}. On the right we provide the dual description of the same theory which is a disjoint union of two good theories each of which flows to an interacting SCFT. These theories were first studied together in \cite{Dey:2014tka}. Our goal in this section is to explain how this theory arises.\\

Although the computation presented in this section can be easily generalized to any $N$, we will focus on $N=4$ for simplicity. In particular, 
the treatment of the bad node remains essentially the same for larger $N$.\\

First, let us analyze the theory on the left in \figref{fig: ex1} in more detail. 
This is indeed a bad theory, since the naive computation of the
scaling dimensions of monopole operators in the IR lead to unitarity violating answers. The monopole operators are labelled
by GNO charges associated with each gauge node -- let us label them as $H_1=(p,-p)$ (for the gauge node with $SO(2N)$ flavor symmetry) 
and $H_2=(n,-n)$. The scaling dimension formula for the monopole operators (which is the UV $U(1)$ R-symmetry charge for a given monopole
operator -- this is the R-symmetry on the Coulomb branch preserved by these BPS operators) reads:
\begin{align}
\Delta_R (N=4)= & - \sum_{i=1,2} \sum_{\alpha \in \Delta^+} |\alpha(H_i)| + \frac{1}{2}\sum_{i\,,\,R} \sum_{\rho \in \Delta(R)} |\rho(H_i)| \\
= & (-2|n| - 2|p|) + 4|p| + |n-p| +|n+p|, \nonumber \\
= & 2|p| - 2|n| + |n-p| +|n+p|.
\end{align}
Evidently, for $p=0$, one encounters unitarity violating scaling dimensions $\Delta_R =0$, for all $n$ (as one expects in an $SU(2)$
theory with two flavors), which implies that the theory is bad. For generic $N$, we have
\begin{align}
\Delta_R (N)= (N-2) |p| - 2|n| + |n-p| +|n+p|\,,
\end{align}
which again gives unitarity violating scaling dimensions for $p=0$, and for any $n$. If $N \leq 2$, there are additional
unitarity violating monopole operators for all $n,p$ such that $n = \pm p$.

We can see that the theory is bad due to the second $SU(2)$ node of the quiver on the left of \figref{fig: ex1}. 
It is obvious that monopole operators charged under the first $SU(2)$ group satisfy the unitarity bound.
The first $SU(2)$ group plays a role of the global symmetry for the second $SU(2)$ group. In other words, the bad sector of this bad theory behaves the same way as an $SU(2)$ theory with two flavors. The Coulomb branch of the latter theory has quaternionic dimension one, and is given by the cone $\mathbb{C}^2/\mathbb{Z}_2$ 
in the IR limit. It was argued in \cite{Seiberg:1996bs} that this moduli space is associated with an interacting SCFT, as opposed to a free $\CN=4$ hypermultiplet 
with a gauged discrete symmetry group. The interacting SCFT can be thought of as the IR fixed point of a $U(1)$ theory with two flavors (the Coulomb branch 
of which is also $\mathbb{C}^2/\mathbb{Z}_2$). This is the reason why the $U(1)$ theory with two flavors appears on the right of \figref{fig: ex1}.

In the rest of the section we shall confirm this claim by studying singular loci of the the Coulomb branch of the bad theory in question. 

\subsection{Abelianized Coulomb branch chiral ring relations}
Let us now analyze the Coulomb branch of the gauge theory on the left in \figref{fig: ex1} as an algebraic variety following \cite{Bullimore:2015lsa} based on the formalism developed in \cite{Assel:2017jgo}. 
Recall that the Coulomb branch of a 3d $\CN=4$ gauge theory with gauge group $G$ is parameterized by VEVs of the vectormultiplets which consist of triplets of real scalars and the dual photon $(\phi_a^1,\phi_a^2,\phi_a^3,\gamma_a)$, where $a=1,\ldots, {\rm rk}G$. These degrees of freedom can be rearranged into complex scalars 
and monopole operators which in the classical description are combinations of the dual photon $\gamma_a$ and the remaining scalar
\begin{equation}\label{class-rel}
\varphi_a = \phi^1_a+i \phi^2_a, \qquad u_a^\pm = \exp\,\pm\left(\frac{2\pi}{g^2_{\text{YM}}} \phi^3_a + i\gamma_a\right)\,,
\end{equation}
where the monopole operators $u_a^\pm$ obey the relations:
\begin{equation}
u_a^+ \, u_a^- =1, \quad \forall a.
\end{equation}
Due to the periodicity in $\gamma$ direction the classical Coulomb branch is a cylinder 
\begin{equation}
\mathcal{C}\simeq (\mathbb{C}\times\mathbb{C}^*)^{\text{rk}G}/\text{Weyl}_G
\end{equation}
modulo the action of the Weyl symmetry. Quantum corrections deform the Coulomb branch and the classical relations \eref{class-rel}. 
It was shown in \cite{Bullimore:2015lsa} that the quantum-corrected Coulomb branch and the relations between monopole operators 
can be described using the ``abelianized" description, i.e. in terms of Weyl-invariant polynomials of $\varphi_a$ and $u_a^\pm$.
In particular, the quantum-corrected Coulomb branch can be realized as a complex algebraic variety in variables $\varphi_a$ and $u_a^\pm$. 
For example, given an SQCD with gauge group $U(N)$ and $N_f$ flavors the classical relations \eref{class-rel} are replaced by the 
following quantum-corrected ones:
\begin{equation}
u_a^+ u_a^- \prod_{b\neq a}^N(\varphi_a-\varphi_b)^2 = \varphi_a^{N_f}\,,\qquad a=1,\dots, N\,.
\label{eq:CbranchBDG}
\end{equation}
The Coulomb branch relations of the non-Abelian theory are obtained by rewriting the above relations in terms of operators in 
the non-Abelian theory, which in turn can be written in terms of Weyl-invariant polynomials of $\varphi_a$ and $u_a^\pm$ (this was 
referred to as the ``abelianization map" in \cite{Bullimore:2015lsa}).\\

Let us now describe the procedure explicitly for a linear quiver gauge theory with unitary gauge groups, where the $i$-th gauge node 
has rank $N_i$, and has $M_i$ fundamental hypermultiplets. The non-Abelian Coulomb branch relations can then be obtained from the following
polynomial equation:
\begin{equation}\label{CRR-gen}
Q_i(z) \widetilde Q_i(z) + U_i^+(z) U_i^-(z) -P_i(z) Q_{i-1}(z) Q_{i+1}(z) =0\,,
\end{equation}
where, for each gauge node $i$ with number of colors $N_i$ and number of flavors $M_i$, we introduce the following generating functions 
\begin{equation}
Q_i(z) = \prod_{a=1}^{N_i} (z-\varphi_{i,a})\,,\, U_i^\pm(z)=\sum^{N_i}_{a=1} u_{i, a}^{\pm} \prod_{b \neq a} (z - \varphi_{i,b}),\quad \widetilde{Q}_i(z) = \prod_{a=1}^{\widetilde N_i} (z-\widetilde\varphi_{i,a})\,,\quad P_i(z)=z^{M_i}\,,
\label{eq:QQtilde}
\end{equation}
where $\widetilde{N}_i = {\rm max} (N_i -2, M_i -N_i + N_{i-1} + N_{i+1})$.
As discussed in \cite{Cremonesi:2013lqa,Bullimore:2015lsa}, the Coulomb branch is generated by a subset of BPS monopole operators with 
magnetic charges $(0,\ldots,0)$ and $(\pm 1, 0,\ldots,0)$. For our purposes it will be more convenient to work with these gauge invariant 
Coulomb branch operators, which are related to the abelian operators in the following fashion:
\begin{align}
&\Phi_{i,n} = \sum_{a_1<\dots a_n} \varphi^{(i)}_{a_1}\cdots\varphi^{(i)}_{a_n}\,, \quad n=1,\ldots, N_i \label{eq:ScalarSym} \\
&V^{\pm}_{i,n}=\sum_{a=1}^{N_i} u_{i, a}^{\pm}\sum_{\substack{b_1<\dots <b_n\\ b_l\neq a}} \varphi^{(i)}_{b_1}\dots\varphi^{(i)}_{b_n}\,.
\end{align}
In terms of the gauge invariant operators, $Q_i(z)$, $\widetilde Q_i(z)$, and $U_i^{\pm}(z)$ is given as:
\begin{align}
& Q_i(z) = \sum^{N_i}_{n_i=0} (-1)^{n_i} \Phi^{(i)}_{n_i} z^{N_i - n_i},  \\
& \widetilde Q_i(z)= \sum^{\widetilde{N}_i}_{n_i=0} (-1)^{n_i} \widetilde \Phi^{(i)}_{n_i} z^{\widetilde{N}_i - n_i}, \\
& U_i^{\pm}(z) = \sum^{N_i -1}_{n_i=0} (-1)^{n_i} V^{\pm}_{i,n} z^{N_i -1-n_i}\,.
%& R_i(z) = \sum^{M_i + N_{i-1} + N_{i+1}}_{k_i=0} (-1)^{k_i} R^{(i)}_{k_i} z^{M_i + N_{i-1} + N_{i+1} - k_i},
\end{align}
where $\widetilde \Phi^{(i)}_{0} =1$. Therefore, the non-Abelian Coulomb branch relations can be read off from \eref{CRR-gen} after 
solving for the auxiliary variables $\widetilde \Phi^{(i)}_{n_i}$.

\subsection{$U(2)$ and $SU(2)$ Theories with Two Flavors}
To begin with, let us analyze the case of a U(2) gauge theory with two flavors, which is a bad theory. We will need this analysis later for the bad quiver from \figref{fig: ex1}.
In terms of \eref{CRR-gen} this theory has a single node whose Coulomb branch relations are generated by
\begin{equation}
Q(z) \widetilde Q(z) + U^+(z) U^-(z) = z^2\,.
\end{equation}
Here $\widetilde Q(z)=\widetilde\Phi_0$ is a constant,  all other polynomials are of degree two. We obtain the following
\begin{align}
\label{eq:Coulomb22}
V_0^+ V_0^- +\widetilde\Phi_0 &=1\,,\cr
V_1^- V_0^+ + V_1^+ V_0^- + \Phi_1\widetilde\Phi_0&=0\,,\\
V_1^- V_1^++ \Phi_2\widetilde\Phi_0&=0\,.\notag
\end{align}
One solves the first equation with respect to $\widetilde\Phi_0$ to obtain the system of equations for the Coulomb branch.
For an $SU(2)$ gauge group, we have 
\begin{equation}
\label{eq:DefinitionPhi}
\Phi_0=1\,,\quad \Phi_1=\varphi_1+(-\varphi_1)=0\,,\quad \Phi_2=\varphi_1(-\varphi_1) = -\varphi_1^2=:-\varphi^2\,,
\end{equation}
since $\varphi_2=-\varphi_1$. 

The Coulomb branch relations for the $SU(2)$ theory can be obtained from the ones for the $U(2)$ theory (as given in \eref{eq:Coulomb22}) 
by imposing the following constraints. Firstly, one needs to impose the vanishing trace condition \eqref{eq:DefinitionPhi} on the complex scalars. 
Secondly, the Coulomb branch operators for the $SU(2)$ theory 
must be invariant under the action of the topological $U(1)_J$ symmetry\footnote{We thank Benjamin Assel for pointing this out to us.}. 
Recall that, in the $U(2)$ theory, the monopole operators $V^{\pm}_k$ have charges $\pm 1$ respectively, while the complex scalars are invariant,
under this symmetry. We will find it convenient to write the chiral ring relations for the $SU(2)$ theory in terms of the following $U(1)_J$-invariant operators $V_{kl}$:
%The above relations for complex scalars VEVs in the $SU(2)$ theory is not the only constraint which we need to impose. 
%We should also regard carefully the dressed monopole operators $V^{\pm}_k$ \eqref{eq:ScalarSym}. Until now we have not discussed the important symmetry which acts on the Coulomb branch operators -- topological $U(1)_J$ symmetry. In the $U(2)$ theory operators $V^{\pm}_k$ have charges $\pm 1$ respectively under this symmetry. When we are imposing vanishing trace condition \eqref{eq:DefinitionPhi} for complex scalars in order to obtain $SU(2)$ operators from $U(2)$ operators we also need to impose this condition for the third scalar components $\phi^3_a$. However, $\phi^3_a$ only appear in semiclassical expressions like \eqref{class-rel}; in quantum regime we need to use the corresponding monopole operators. Thus, in order to describe the Coulomb branch of the $SU(2)$ theory using $U(2)$ operators we need to \textit{gauge} the topological $U(1)_J$ symmetry\footnote{We thank Benjamin Assel for pointing this out to us.}. Once the topological symmetry is gauged operators $V^+_k$ and $V^-_k$ are not gauge invariant any longer, therefore we need to use their composites, which are gauge invariant. We can define the following operators $V_{kl}$, neutral under $U(1)_J$ which at weak coupling can be represented as follows
\begin{equation}
V_{kl} := V^+_k V^-_l\,.
\label{eq:U1JInv}
\end{equation}

From the above expression it is clear that operators $V_{kl}$ are not independent. Indeed, one needs to impose
\begin{equation}
\text{det}_{a,b} V_{ab} = 0\,,
\label{eq:MinorRelation}
\end{equation}
for any $2\times 2$ minor of the matrix $V_{kl}$.

We are now ready to discuss the Coulomb branch of the $SU(2)$ theory with two flavors. Since in the $U(2)$ theory we only have $V_0^\pm$ and $V_1^\pm$ monopole operators then \eqref{eq:MinorRelation} gives a single relation:
\begin{equation}
R_1:=V_{00} V_{11} - V_{01} V_{10}=0\,,
\label{eq:Det22}
\end{equation}
while relations \eqref{eq:Coulomb22} will read as follows after eliminating $\widetilde\Phi_0$:
\begin{align}
%\label{eq:Coulomb222}
& R_2:= V_{10} + V_{01} =0\,,\label{eq:Coulomb222a}\\
& R_3:= V_{11}+ \Phi_{2}(1-V_{00})=0\, \label{eq:Coulomb222b}.
\end{align}

%Combining \eqref{eq:Det22} and \eqref{eq:Coulomb222} together we get
%\begin{equation}
%\Phi_{2}V_{00}(1-V_{00})-V_{01}^2 =0\,,
%\label{eq:SU22C}
%\end{equation}
%which is the desired relation for the Coulomb branch. 
%Now we would like to discuss its singularities. 
%are given by vanishing of the 
%Jacobian of the left hand side of \eqref{eq:SU22C}
Note that we have three relations $\{R_l\}$ given by \eref{eq:Det22}-\eref{eq:Coulomb222b} in terms of five 
variables $\{\CO_i\}=(V_{00}, V_{01}, V_{10}, V_{11}, \Phi_2)$. The singular loci of the moduli space will 
correspond to points where the Jacobian matrix $J^i_{\,\,l} =\frac{\partial R_l}{\partial \CO_i}$ has rank less
than three. Explicitly, the matrix is given as:
\begin{equation}
J^i_{\,\,l} = \begin{pmatrix} V_{11} & 0 & - \Phi_2 \\ -V_{10} & 1 & 0 \\  -V_{01} & 1 & 0 \\ -V_{00} & 0 & 1 \\ 0 & 0 & (1-V_{00}) \end{pmatrix}.
\end{equation}
%\begin{align}
%V_{00}(1-V_{00})&=0\,,\cr
%\Phi_{2}(1-2V_{00})&=0\,,\cr
%V_{01}&=0\,.
%\label{eq:JacSu22}
%\end{align}
Now, we would like to study the singular loci of the Coulomb branch. The analysis can be simplified by eliminating the variables 
$V_{11}$ and $V_{10}$ using the relations \eref{eq:Det22} and \eref{eq:Coulomb222a}, so that we have three variables $(V_{00}, V_{01}, \Phi_2)$ 
and a single relation:
\begin{equation}
R:=\Phi_{2}V_{00}(1-V_{00})-V_{01}^2 =0\,.
\label{eq:SU22C}
\end{equation}
The singular loci, which correspond to vanishing of the Jacobian $J^i = \frac{\partial R}{\partial \CO_i}$, where $i=1,2,3$ (i.e. a rank zero Jacobian),
is given by the following set of equations:
\begin{align}
V_{00}(1-V_{00})&=0\,,\cr
\Phi_{2}(1-2V_{00})&=0\,,\cr
V_{01}&=0\,.
\label{eq:JacSu22}
\end{align}
Therefore, the singular loci on the Coulomb branch consists of two distinct points, corresponding to the two solutions of the above equations:
\begin{align}
&(1):\quad V_{00} = 0,\, \Phi_{2}=0,\,V_{01}=0,\,\cr 
&(2):\quad V_{00} = 1,\, \Phi_{2}=0,\, V_{01}=0\,.
\end{align} 
In the vicinity of the above singularities,  we consider small fluctuations of the fields and parametrize them as follows:
\begin{align}
&(1):\quad V_{00} = v, \,\Phi_{2}=u, \,V_{01}=w\,,\cr 
&(2):\quad V_{00} = 1 - v, \,\Phi_{2}=u, \,V_{01}=w\,,
\end{align}
where $u,v$ and $w$ are small. In both cases, the relation \eref{eq:SU22C} (or the relations \eref{eq:Det22}-\eref{eq:Coulomb222b}) gives the equation for an $A_1$ singularity to the leading order in fluctuations:
\begin{equation}
uv=w^2\,.
\end{equation}

Therefore we conclude that the Coulomb branch of a 3d $\mathcal{N}=4$ $SU(2)$ theory with two flavors has two distinct singular points where 
the moduli space is locally isomorphic to $\mathbb{C}^2/\mathbb{Z}_2$ as an algebraic variety \footnote{The $SU(2)$ theory is a representative of a family of $Sp(k)$ theories with $SO(2N)$ global symmetry for $k=1$. Bad theories of this class will be analyzed in an upcoming paper \cite{Assel:2018a} by the authors of \cite{Assel:2017jgo}.}.

%$\mathbb{C}^2/\mathbb{Z}_2$ singularities each of which describes $U(1)$ theory with two flavors

\subsection{$SU(2)$ Theory with Four Flavors}
Let us now describe Coulomb branch of a good theory -- $SU(2)$ with four flavors, which will be used in what follows. 
Generating relation \eref{CRR-gen} 
\begin{equation}
Q(z) \widetilde Q(z) + U^+(z) U^-(z) = z^4\,,
\end{equation}
after eliminating auxiliary fields from $\widetilde Q(z)$ and switching to $SU(2)$ variables, which we described above, we get the following
\begin{align}
\label{eq:Coulomb2244}
V'_{00} V'_{11} - V'_{01} V'_{10}&=0\,,\cr
V'_{10} + V'_{01} &=0\,,\cr
V'_{11}- \Phi'_2(\Phi'_2+V'_{00})&=0\,,
\end{align}
where we put primes to distinguish these operators from those of the previous subsection. Using the second relation 
to eliminate the variable $V'_{10}$, we have the reduced set of equations: 
\begin{align}
\label{eq:Coulomb224}
V'_{00} V'_{11} + (V'_{01})^2&=0\,,\cr
V'_{11}- \Phi'_2(\Phi'_2+V'_{00})&=0\,.
\end{align}

\subsection{Bad linear quiver}\label{Sec:GoodBad}
Finally we can proceed to analyze the linear quiver on the left hand side of figure \figref{fig: ex1}. 
For the right node of the quiver \eqref{CRR-gen} reads\footnote{To avoid index cluttering we shall use primes for operators charged under the left $SU(2)$ node of the quiver.}
\begin{equation}
 Q(z) \widetilde Q(z) + U^{+}(z) U^{-}(z) =Q'(z)\,.
\end{equation}
and yields the following equations 
\begin{align}
\label{eq:Coulomb22x}
V_0^+ V_0^- +\widetilde\Phi_0 &=1\,,\cr
V_1^- V_0^+ + V_1^+ V_0^- + \Phi_1\widetilde\Phi_0&=0\,,\\
V_1^- V_1^++ \Phi_2\widetilde\Phi_0&=\Phi'_2\,,\notag
\end{align}
where $\widetilde\Phi_0$ is auxiliary variable, which is a part of the abelianization procedure and $\Phi'_2$ is the scalar operator of the form \eqref{eq:ScalarSym} for the left $SU(2)$ node of the quiver.  One solves the first equation with respect to $\widetilde\Phi_0$ to obtain the system of equations for the Coulomb branch.

Since both gauge groups are of the quiver are $SU(2)$ then $\varphi_2=-\varphi_1$ and $\varphi'_2=-\varphi'_1$, therefore the scalar operators read as follows
\begin{equation}
\label{eq:DefinitionPhix}
\Phi_0=1\,,\quad \Phi_1=\varphi_1+(-\varphi_1)=0\,,\quad \Phi_2=\varphi_1(-\varphi_1) = -\varphi_1^2=:-\varphi^2\,,\quad \Phi'_2= -\varphi'^2\,,
\end{equation}
By taking the above formulae into account and switching to $U(1)_J$-invariant variables \eqref{eq:U1JInv} we get the following relations 
from \eqref{eq:Coulomb22x} (after eliminating $V_{10}$ and $\widetilde\Phi_0$) :
\begin{align}
\label{eq:Coulomb222z}
V_{00}V_{11} + V_{01}^2 &=0\,,\cr
V_{11}+ \Phi_2(1-V_{00})&=\Phi'_2\,,
\end{align}
which, after eliminating $V_{10}$, give a single relation:
\begin{equation}
\Phi'_2 V_{00}-\Phi_2 V_{00} (1-V_{00})+V_{01}^2 =0\,.
\label{eq:Phi2bad}
\end{equation}

We can now derive the equations for the left (good) node of the bad quiver in \figref{fig: ex1}.
Equations \eqref{CRR-gen} read (here primes correspond to the operators of the good node)
\begin{equation}
 Q'(z) \widetilde Q'(z) + U^{'+}(z) U^{'-}(z) =z^4 Q(z)\,.
\end{equation}
This relation generates six nontrivial equations which describe scalars and monopole operators for the first node of the quiver. 
It is easy to see that only in one of those equations (coming from $z^4$ term) there will be a contribution from the bad node in terms of operator $\Phi_2$. One now needs to use the remaining five equations to solve for tilded variables. Having done these calculations we get the following equations in gauge invariant variables
\begin{align}
\label{eq:Coulomb111}
V'_{11} V'_{00} + (V'_{01})^2 &=0\,,\cr
V'_{11}-\Phi'_2 V'_{00}&=X^{'6}\,,
\end{align}
where we define $X'$ as
\begin{equation}
X^{'6}=-X^2 \Phi^{'2}_2 :=\Phi^{'2}_2(\Phi_2-\Phi'_2)\,, \quad\quad X^2 =(\Phi'_2-\Phi_2) .
\end{equation}
We can now redefine all primed monopole operators from \eqref{eq:Coulomb111} as follows
\begin{equation}
V'_{ab} = \widetilde V'_{ab} X^2\,.
\end{equation}
Then these equations will read
\begin{align}
\label{eq:Coulomb111eq}
\widetilde V'_{11}  \widetilde V'_{00} + \left(\widetilde V'_{01}\right)^2 &=0\,,\cr
\widetilde V'_{11}-\Phi'_2 \widetilde V'_{00}&=\Phi^{'2}_2\,,
\end{align}
which coincides with \eqref{eq:Coulomb224} upon identification $\widetilde V'_{ab}$ with $V'_{ab}$. Recall that system \eqref{eq:Coulomb224} describes the Coulomb branch of $SU(2)$ theory with four flavors.
The above two equations can be combined to one relation
\begin{equation}
\Phi'_2  (\widetilde V'_{00})^2 + (\Phi'_2)^2 \widetilde V'_{00} + \left(\widetilde V'_{01}\right)^2 =0\,.
\label{eq:QuiverVarSec}
\end{equation}

In order to analyze singular behavior of equations \eqref{eq:Phi2bad} and \eqref{eq:QuiverVarSec} we again look at the locus where the corresponding Jacobian vanishes. There are two algebraic relations for six variables $V_{00},\widetilde V'_{00}, \Phi_2,\Phi'_2, V_{01},\widetilde V'_{01}$, so the Jacobian is a $2\times 6$ matrix. All its $2\times 2$ minors must vanish in order to define the singular locus. After a simple computation we obtain two singular loci for unprimed variables.
\begin{align}
((\Phi'_2)^2+2\Phi'_2 \widetilde V'_{00})(\Phi'_2-\Phi_2(1-2V_{00})&=0\,,\cr
V_{00}(1-V_{00})((\widetilde V'_{00})^2+2\Phi'_2 \widetilde V'_{00})&=0\,,\cr
\widetilde V'_{01}V_{00}&=0\,,\cr
V_{00}((\widetilde V'_{00})^2+2\Phi'_2 \widetilde V'_{00})&=0\,,\cr
V_{01}((\Phi'_2)^2+2\Phi'_2V_{00})&=0\,.
\end{align}
The first singular sublocus is
\begin{equation}
V_{00}=0,\, \Phi_2'=\Phi_2,\, V_{01} = 0\,, 
\label{eq:locusline}
\end{equation}
and the other primed variables are only constrained by \eqref{eq:QuiverVarSec}. 
Small fluctuations around the above singular point
\begin{equation}
V_{00} = v,\, \Phi_2 = u,\, \Phi_2' = u',\, V_{01} = w
\label{eq:Assignments}
\end{equation}
lead to the equation for $A_1$ singularity
\begin{equation}
xv = w^2, \quad \quad x= u-u'.
\label{eq:A1singquiv}
\end{equation}

The second singular sublocus reads
\begin{equation}
V_{00}=1,\, \Phi_2'=-\Phi_2,\, V_{01} = 0\,, \widetilde V'_{01}=0\,,
(\widetilde V'_{00})^2+2\Phi'_2 \widetilde V'_{00}=0\,,
\end{equation}
which after substituting to \eqref{eq:Phi2bad} gives
\begin{equation}
V_{00}=1,\, \Phi_2'=\Phi_2=0,\, V_{01} = 0\,.
\label{eq:V2sing}
\end{equation}
In other words, only fluctuations around the above locus give the equation for the $A_1$ singularity, instead of having an entire complex line in \eqref{eq:locusline}. Other primed variables also obey \eqref{eq:QuiverVarSec}. Analogously to the previous case we conclude that small fluctuations around that locus describe the $A_1$ singularity.

Note that there is a difference between the above two cases. In \eqref{eq:locusline} the Coulomb branch of the given quiver theory is a direct product of the $A_1$ singularity and the Coulomb branch of the $SU(2)$ theory with four flavors described by \eqref{eq:Coulomb224}. However, singularity \eqref{eq:V2sing} is of higher codimension and in its vicinity the Coulomb branch decomposes into the $A_1$ singularity and a singular sublocus of the Coulomb branch of the $SU(2)$ theory with four flavors.

\subsection{Matching UV and IR R-symmetries}
Having a good description of the bad theory allows us to see the explicit relationship between the UV Coulomb branch R-symmetry $SU(2)_C$ and the IR R-symmetry based on the above calculation. Indeed, since for good theories R-symmetries do not change along the renormalization group flow, we can use the theory from the right in \figref{fig: ex1} to read off the IR R-symmetry of the sought bad theory. 

According to \eqref{eq:Coulomb222a}-\eqref{eq:Coulomb222b} and \eqref{eq:A1singquiv} the UV operators $V_{00}$ are neutral under $U(1)_C\subset SU(2)_C$. In other words $V_{00}$ violates the unitarity bound and in the good description of the theory they will acquire some positive R-charge. 

Let us look at \eqref{eq:A1singquiv} more closely
\begin{equation}
xv = w^2\,,
\end{equation}
where $x=u+u'$. According to \eqref{eq:Assignments} the UV R-charges are
\begin{equation}
[x]_{\text{UV}}=2,\quad [v]_{\text{UV}} = 0,\quad [w]_{\text{UV}}=1\,.
\end{equation}
In the infrared $x$ and $v$ play the role of monopole operators for the $U(1)$ theory with 2 flavors, while $w$ is the VEV for the complex scalar. We can easily determine R-charges of these operators, which for the quiver UV theory we have started with will play the role of the IR R-charges
\begin{equation}
[x]_{\text{IR}}=1,\quad [v]_{\text{IR}} = 1,\quad [w]_{\text{IR}}=1\,.
\end{equation}
Additionally the $U(1)$ theory has a $U(1)_J$ topological symmetry, which for this theory is enhanced to $SU(2)_J$. This symmetry mixed with the IR R-symmetry should give the UV R-symmetry
\begin{equation}
SU(2)_C = \text{diag}(SU(2)_J\times SU(2)_{IR})\,.
\end{equation}
Based on the above considerations the assignments of the $U(1)_J\subset SU(2)_J$ are the following
\begin{equation}
[x]_{J}=1,\quad [v]_{J} = -1,\quad [w]_{J}=0\,.
\end{equation}

Note that according to the assignments of the IR R-charges all generators $x,v$ and $w$ can be thought of as a top components of the $SU(2)_\text{IR}$ triplets. Their complex conjugates $\bar x, \bar v$ and $\bar w$ are the bottom components. It is still to be determined what the middle components with zero R-charge are. Nevertheless this observation is in one-to-one correspondence with Bashkirov's classification of 3d $\mathcal{N}=4$ IR SCFTs. Indeed, this example corresponds to the presence of flavor supercurrent multiplet $\mathcal{F}$ which carries triplet $\textbf{3}$ of $SU(2)_{IR}$. We have identified three such triplets.

\section{Partition Function Analysis}\label{PFA}
In this section, we present a derivation of the good dual in \figref{fig: ex1} using partition function on a three-sphere. 
It is well-known that the three-sphere partition function diverges for bad theories.
However, it was shown in \cite{Willett:2011gp, Yaakov:2013fza} that the three-sphere partition function of a bad theory with $U(N_c)$ gauge group
and $N_c \leq N_f < 2N_c -1$ flavors can be appropriately regularized by turning on generic R-symmetry charges. Explicitly,
\begin{equation}
\begin{split}
&\mathcal{Z}_{N_{c},N_f}^{\rm reg}\left(\{m_{a}\};\eta\right):=I_{N_{c},(2,2)}^{N_{f}-N_{c}}\left(\mu_a=\frac{\omega}{2}-m_{a},\nu_a=\frac{\omega}{2}+m_{a},-2\eta|\omega_{1},\omega_{2}\right),\\
&I_{n,(2,2)}^{m}(\mu_a,\nu_a,\lambda |\omega_{1},\omega_{2})=\frac{1}{(-\omega_{1}\omega_{2})^{\frac{n}{2}}n!}\int\limits_{C^{n}} \prod_{j=1}^{n} ds^{j} \prod_{1\leq j<k\leq n}\frac{1}{\Gamma_{h}(\pm(s^{j}-s^{k}))}\\
&\times \prod_{j=1}^{n}e^{\frac{\pi i\lambda s^{j}}{\omega_{1}\omega_{2}}}\prod_{a=1}^{n+m}\Gamma_{h}(\mu_{a}-s^{j})\Gamma_{h}(\nu_{a}+s^{j})\,,\qquad \qquad \omega=\frac{\omega_{1}+\omega_{2}}{2},
\end{split}
\end{equation}
where $\Gamma_h$ denotes a hyperbolic gamma function, the parameters $\{m_a\}$ are real mass parameters living in the Cartan subalgebra of 
$U(N_f)$, and $\eta$ is the FI parameter. $C^n$ is generically a contour on the complex plane which will be taken to be the real line for our purposes. The complex numbers $(\omega_1, \omega_2)$ take values $(\I, \I)$ on a round three-sphere and $(\I b, \I/b)$ on a squashed three-sphere, while $\I \omega/2$ is associated with the IR dimension of a chiral multiplet (the canonical dimension being 1/2). 
We refer the reader to \cite{ELHFn,Willett:2011gp} for further details.\\

Starting from the partition function of an ugly theory and integrating out one flavor at a time,
the regularized partition function of the aforementioned bad theory can be written in terms of the partition function of a good theory, i.e. a
$U(N_f - N_c)$ theory with $N_f$ flavors, and that of $2N_c-N_f$ free hypermultiplets \cite{Yaakov:2013fza}, i.e.
\begin{equation} \label{Y-NcNf}
\begin{split}
\mathcal{Z}_{N_{c},N_f}^{\rm reg}\left(\{m_{a}\}; \eta \right)
= \mathcal{Z}_{N_{f}-N_c,N_f} \left(\{m_{a}\}; -\eta \right)\cdot \mathcal{Z}^{\rm b.g.}_{\rm FI}(\{m_{a}\},\eta)\cdot \mathcal{Z}_{\rm hyper}\left(\eta_u\right)\cdot \prod^{2N_c - N_f -1}_{j=1} \mathcal{Z}_{\rm hyper} \left(\eta_j\right),
\end{split}
\end{equation}
where the contributions for the hypermultiplets $\mathcal{Z}_{\rm hyper}$ and background Fayet-Iliopoulos $\mathcal{Z}^{\rm b.g.}_{\rm FI}$ terms are given as
\begin{align}
&\mathcal{Z}_{\rm hyper}\left(\eta\right)= \Gamma_h\left(\frac{\omega}{2} + \eta\right)\Gamma_h\left(\frac{\omega}{2}- \eta\right),\\
&\mathcal{Z}^{\rm b.g.}_{\rm FI}(\{m_{a}\},\eta)= \exp{\Big(\frac{2\I \eta}{\omega_1 \omega_2}\,\sum_a m_a\Big)}\,,
\end{align}
and the parameters $\eta_u, \eta_j$ are:
\begin{align}
& \eta_u = \eta - (2N_c - N_f -1)\frac{\omega}{2},  \quad \eta_j= -\eta + (2N_c -N_f +1 -2j)\frac{\omega}{2}.
\end{align}

The expression for the dual partition function of an $SU(N_c)$ with $N_f$ flavors can be obtained by integrating both sides 
of \eref{Y-NcNf} w.r.t. the FI parameter $\eta$, although the resultant partition function does not factorize as cleanly as the 
$U(N_c)$ case. However, in the case of $N_c=N_f$, the equation \eref{Y-NcNf} simplifies further as the RHS of the equation 
reduces to the contribution of $N_f$ free hypermultiplets, i.e.
\begin{align}
\mathcal{Z}_{N_{c},N_c}^{\rm reg}\left(\{m_{a}\}; \eta \right) = \mathcal{Z}^{\rm b.g.}_{\rm FI}(\{m_{a}\},\eta)\cdot \mathcal{Z}_{\rm hyper}\left(\eta_u\right)\cdot 
\prod^{N_c -1}_{j=1} \mathcal{Z}_{\rm hyper}\left(\eta_j\right).
\end{align}
In this case, the dual partition function of $SU(N_c)$ theory with $N_c$ flavors can be written as
\begin{align}
& \mathcal{Z}_{SU(N_{c}),N_c}^{\rm reg}\left(\{m_{a}\}\right) = \int \de\eta \, \mathcal{Z}^{\rm b.g.}_{\rm FI}(\{m_{a}\},\eta)\cdot \mathcal{Z}_{\rm hyper}\left(\eta_u\right)\cdot 
\prod^{N_c -1}_{j=1} \mathcal{Z}_{\rm hyper}\left(\eta_j\right), \\
& \eta_u = \eta - (N_c -1)\frac{\omega}{2},  \quad \eta_j= -\eta + (N_c  +1 -2j)\frac{\omega}{2}. \nonumber
\end{align}
For the specific case of $N_c=N_f=2$, we have:
\begin{align}
\mathcal{Z}_{SU(2), 2}^{\rm reg}\left(\{m_{a}\}\right) = & \int \de\eta \, \mathcal{Z}^{\rm b.g.}_{\rm FI}(\{m_{a}\},\eta)\cdot \mathcal{Z}_{\rm hyper}\left(\eta - \frac{\omega}{2}\right)\cdot \mathcal{Z}_{\rm hyper}\left(-\eta + \frac{\omega}{2}\right) \label{SU2-U1}\\\label{eq:Zsplit}
=& \mathcal{Z}_{U(1), N_f=2} \left(\left\{\widetilde{m}_\ell = -\frac{\omega}{2}\right\}, \widetilde{\eta}=\sum_a m_a\right),
%=& \mathcal{Z}_{U(1), N_f=2} (\{\widetilde{m}_\ell =0\}, \widetilde{\eta}=\sum_a m_a), 
\end{align}
where $\{\widetilde{m}_\ell\}$ and $\widetilde{\eta}$ are the mass and FI parameters of the $U(1)$ theory with $N_f=2$.\\

One can now find a dual for the bad quiver in \figref{fig: ex1} in the following fashion. We can think of constructing this quiver 
by starting out with an $SU(2)$ gauge theory with two flavors and then gauging the flavor symmetry node as an $SU(2)$, followed 
by adding $N$ fundamental hypers to the new gauge node. The strategy is to write the regularized partition function of the bad quiver,
and insert the dual of the bad $SU(2)$ node in the expression, as discussed above. The dual of the bad quiver can then be read 
off from the partition function after some simple manipulation. Explicitly, one can write the regularized partition function as:
\begin{equation}\label{badquiver-1}
\begin{split}
&\mathcal{Z}^{ {\rm reg}}_{\rm bad\,quiver}= \int \prod_a \de m_a \, \delta(m_1 + m_2) \, \mathcal{Z}^{\rm 1-loop,\, reg}_{U(2),\, N_f=N} (\{m_a\},\{M_i\}) \,\cdot \mathcal{Z}^{ {\rm reg}}_{SU(2),\, N_f=2} (\{m_a\})\\
=& \int \prod_a \de m_a \, \delta(m_1 + m_2) \, \mathcal{Z}^{\rm 1-loop,\, reg}_{U(2),\, N_f=N} (\{m_a\}, \{M_i\}) \, \cdot \mathcal{Z}^{\rm reg}_{U(1), N_f=2} \left(\{\widetilde{m}_\ell = -\frac{\omega}{2}\}, \widetilde{\eta}=\sum_a m_a\right),
\end{split}
\end{equation}
where $\mathcal{Z}^{\rm 1-loop,\, reg}$ is the 1-loop contribution of the good node to the quiver partition function 
and $N$ fundamental flavors, where the $N$ mass parameters $\{M_i\}$ live in the Cartan of the $SO(2N)$ flavor symmetry group.
Note that the mass parameters $\{m_a\}$ associated with the bad $SU(2)$ node have been identified with the chemical potentials 
of the good $SU(2)$ gauge node.

Finally, implementing the delta-function constraint on the second term in \eref{badquiver-1}, we have a factorization of the partition function 
into two terms:
\begin{equation}\label{eq:ZsplitFinal}
\begin{split}
\mathcal{Z}^{ {\rm reg}}_{\rm bad\,quiver}
=& \int \prod_a \de m_a \, \delta(m_1 + m_2) \, \mathcal{Z}^{\rm 1-loop,\, reg}_{U(2),\, N_f=N} (\{m_a\}, \{M_i\}) \, \times \mathcal{Z}^{\rm reg}_{U(1), N_f=2} (\{\widetilde{m}_\ell =0\}, \widetilde{\eta}=0) \\
=& \mathcal{Z}^{\rm reg}_{SU(2), N_f=N}(\{M_i\}) \times \mathcal{Z}^{\rm reg}_{U(1), N_f=2} (\{\widetilde{m}_\ell = 0 \}, \widetilde{\eta}=0),
\end{split}
\end{equation}
where for the first equality we have also set the masses $\{\widetilde{m}_\ell =0\}$ by shifting the integration variable $\eta$ in \eref{SU2-U1}
\footnote{One might worry if this shift of the integration variable, which essentially amounts to a contour shift along the imaginary axis, requires 
adding/subtracting residues of poles which are being excluded/included as a result. In the present case, these residues are proportional to 
$\sum_a m_a$ and therefore vanish on implementing the delta function constraint. This is most easily checked in the case of a round three-sphere.}. 
The above equation suggests that the duality in \figref{fig: ex1} only holds when the masses and the FI parameters for the $U(1)$ theory 
are all set to zero.

Our analysis suggest that factorization of the partition function in the above calculation, as well as factorization of the moduli spaces in \secref{Sec:CoulmbBranch}, is only possible for quivers with $SU(2)$ nodes. Indeed, only for $N_c=N_f=2$ in \eqref{eq:Zsplit} we get such symmetric form of the contributions from the hypermultiplets which leads to splitting of the partition function in \eqref{eq:ZsplitFinal}. 

\paragraph{Generalizations.}
We expect to find more examples of 3d $\CN=4$ theories whose IR SCFTs represent themselves a union of interacting sectors. An obvious generalization of the main example in the present paper \figref{fig:QuiverflavoredD44} is to take a `star-shaped' quiver with $SU(2)$ node with $SO(2N)$ framing ($N\geq 4$) in the middle and connect it with $n$ $SU(2)$ nodes.  Each of these $SU(2)$ nodes yields a bad sector. Then the proposed good description will have $n$ decoupled sectors each of which describes an $A_1$ singularity together with the $SU(2)$ theory with $SO(2N)$ global symmetry. In the future work we plan on studying such new examples.

\section*{Acknowledgements}
We would like to thank Denis Bashkirov, Benjamin Assel and Stefano Cremonesi for helpful suggestions and discussions. AD and PK would like to thank Simons Center for Geometry and Physics and the organizers of the 2017 Summer Workshop, where this project has started. PK acknowledges support of IH\'ES and funding from the European
Research Council (ERC) under the European Union's Horizon 2020
research and innovation program (QUASIFT grant agreement 677368). Additionally PK would like to thank Kavli Institute for Theoretical Physics, Tsinghua University, Beijing, and Tsinghua Sanya International Mathematics Forum where part of this work was done. PK's research was also supported in part by the National Science Foundation under Grant No. NSF PHY-1125915.

\bibliography{cpn1-1}
\bibliographystyle{JHEP}

\end{document}